\documentclass[final]{aipproc}

\layoutstyle{6x9}

\usepackage{graphicx,epsfig}
\usepackage{psfig}
\usepackage{amssymb}
\usepackage{wrapfig}
\RequirePackage{lineno}
\begin{document}




\title{Inclusive Dijet Cross Sections in Neutral Current Deep Inelastic Scattering
and Photoproduction at HERA}

\author{Oleg Kuprash}{
address={\vskip-1mm(for the ZEUS Collaboration) \\
Deutsches Elektronen-Synchrotron, Notkestra\ss{}e 85, 22607 Hamburg, Germany}
,altaddress={Taras Shevchenko National University of Kyiv}
}




\begin{abstract}
Recent results from the \emph{ep} collier HERA are presented. Inclusive dijet
cross sections have been measured in neutral current deep inelastic 
scattering, for virtualities of the exchanged boson in the range 
$125 < Q^{2} < 20\,000\; \mbox{GeV}^{2}$ and in photoproduction, $Q^{2} \sim 0\mbox{ GeV}^{2}$. The 
measurements are compared to perturbative QCD calculations at next-to-leading 
order.
\end{abstract}



\maketitle
\section{Introduction}

At HERA, two kinematic regions can be distinguished: in
deep inelastic scattering (DIS) the electron interacts with a parton from the 
proton via the exchange of a virtual boson with large virtuality, 
$Q^{2}$. In contrast, in photoproduction the exchanged photon is quasi-real 
and the electron escapes the detector through the beam pipe. The virtuality of 
the  exchanged boson in DIS is $Q^{2} \gtrsim 1\;\mbox{GeV}^{2}$, whereas in  
photoproduction the exchanged boson is almost on its mass shell, and 
$Q^{2} \lesssim 1\;\mbox{GeV}^{2}$ holds.

The measurements of jet production are a well established tool for stringent 
tests of quantum chromodynamics (QCD) and have been performed at HERA for 
many jet observables. The production of jets allows a direct measurement
of the strong coupling constant, $\alpha_{s}$, and photon (in photoproduction) 
and proton parton density functions (PDFs) can be extracted. 

\begin{figure}[h]
\parbox[c]{5.cm}{\epsfysize=20mm
\epsffile{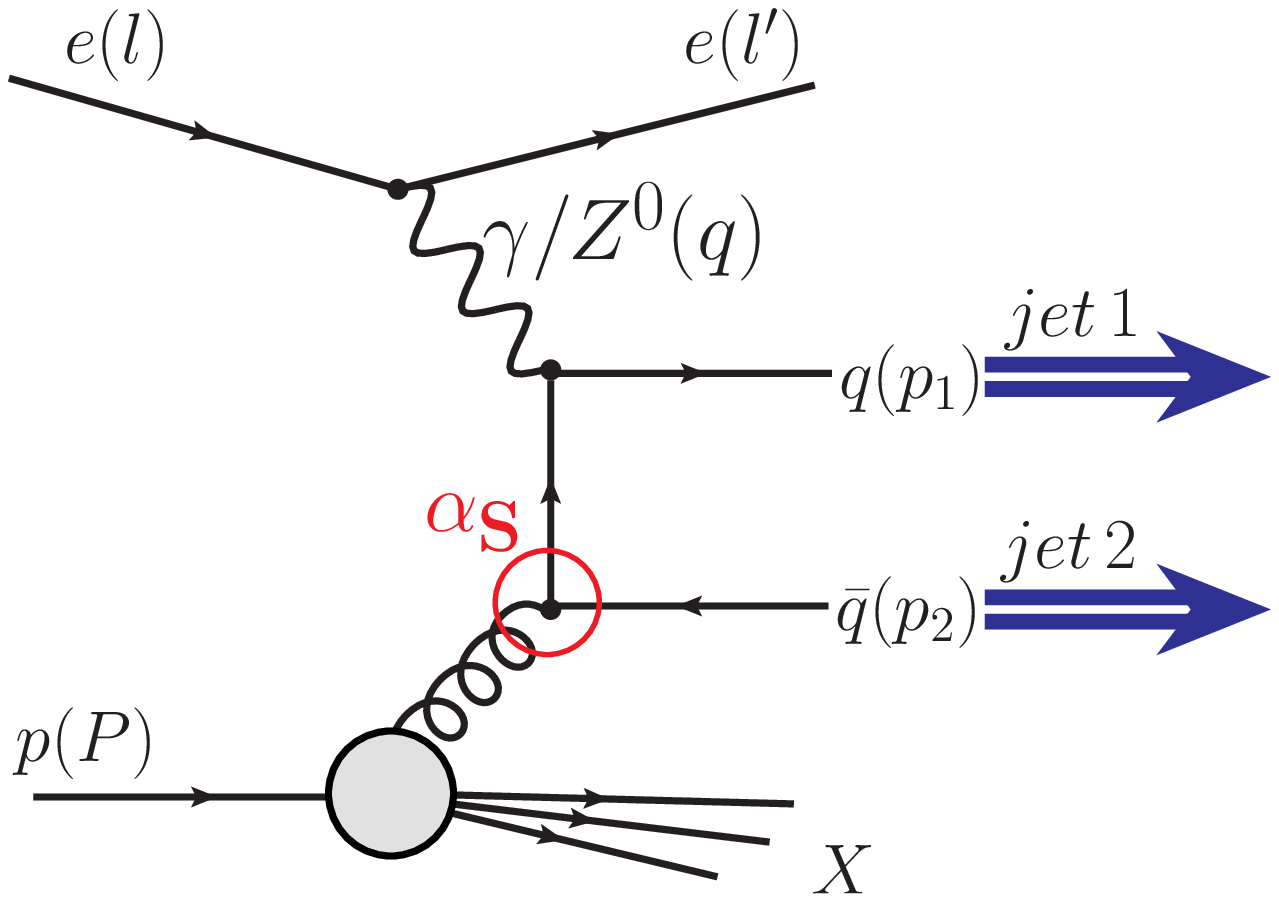}} \hfill~\parbox[c]{5.cm}{\epsfysize=20mm
\epsffile{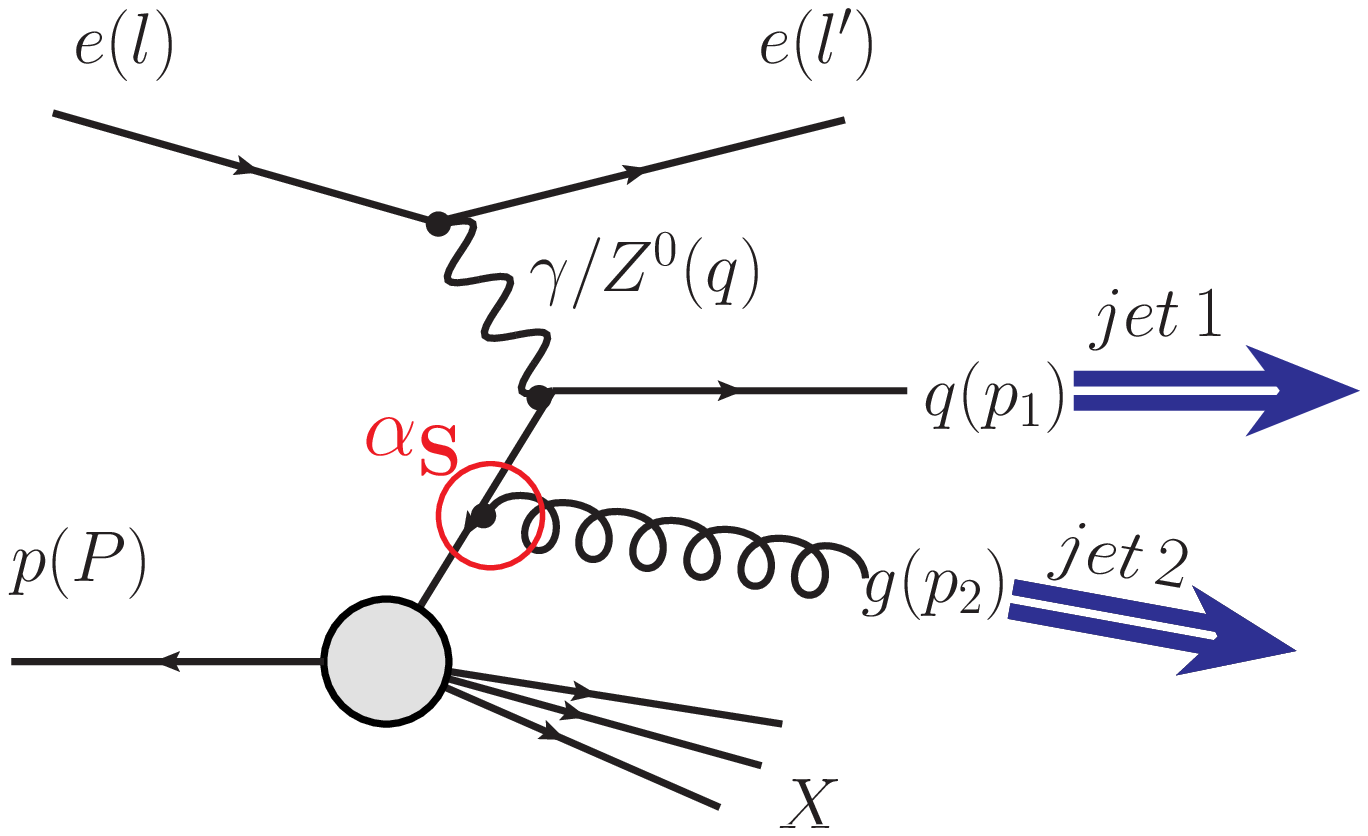}} \hfill~\parbox[c]{5.cm}{\epsfysize=20mm
\epsffile{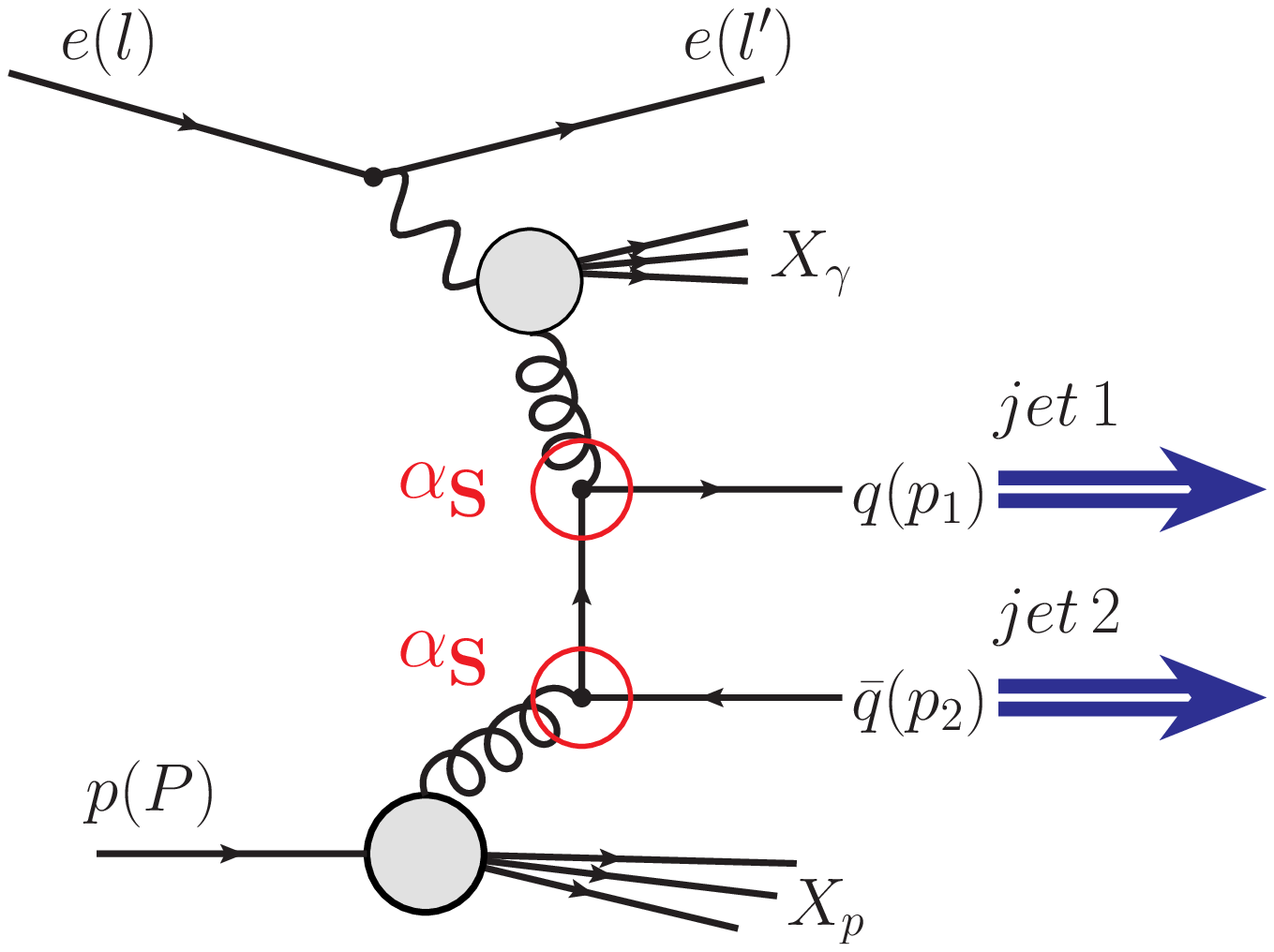}}
\caption{ Boson-Gluon Fusion (left);
QCD Compton scattering (middle);  
LO diagram for resolved photoproduction (right). }
\label{fig:DIS_dijets_diagrams}
\end{figure}

In DIS, two processes contribute to the production of two jets in leading-order 
(LO) $\alpha_{s}$: boson-gluon fusion (Fig. 
\ref{fig:DIS_dijets_diagrams}, left) and QCD Compton scattering (Fig. 
\ref{fig:DIS_dijets_diagrams}, middle). For the dijet measurement in DIS, 
the Breit reference frame was used, since it 
provides a maximal separation between the hard jets and the beam fragmentation 
products. In this frame the exchanged boson collides head-on with the parton.
Therefore, transverse energies are an indicator for the occurrence of strong 
processes. Dijet measurements in DIS have been performed at large virtualities
$Q^{2}$, where both the theoretical and experimental 
uncertainties are relatively small. Jet data have been 
included in the ZEUS-JETS \cite{zeus-jets} fit of the proton PDFs, which significantly reduced 
the uncertainty on the gluon density in the regions of medium and high $x$.

In the photoproduction regime two types of processes contribute to dijet
production in LO $\alpha_{s}$: in direct photon processes the 
photon interacts with a parton as a point-like object whereas in resolved photon processes 
(Fig. \ref{fig:DIS_dijets_diagrams}, right) the photon acts as 
a source of partons, one of which interacts with the parton from the proton. 
Thereby, in addition, in photoproduction the measurement is sensitive to the
photon  PDFs.

In both DIS and photoproduction, jets were reconstructed with the $k_{T}$ 
cluster algorithm \cite{ktclus} in the longitudinally invariant inclusive mode 
\cite{inclusivemode} using the
smallest calorimeter units called cells. The jet search in DIS was performed in the 
Breit reference frame, whereas in photoproduction it was performed in the 
laboratory frame. In order to take into account detector effects, LO Monte Carlo (MC)
samples were used, which utilise different approaches for the parton
cascade. The next-to-leading order (NLO) QCD predictions were corrected using
these MCs to take into account hadronisation and electro-weak effects.

In this report, recent measurements of inclusive dijet production in neutral 
current (NC) DIS \cite{Joerg} and photoproduction \cite{Inna} performed with 
the ZEUS  detector are presented.

The measurements presented here correspond to integrated luminosities of 
189 pb$^{-1}$ for the photoproduction and of 374 pb$^{-1}$ for the NC DIS
analysis, respectively.

\section{Inclusive dijets in neutral current DIS}

The phase space of the measurement was defined by 
$125 < Q^{2} < 20\,000\;\mbox{GeV}^{2}$ and $0.2 < y < 0.6$, where $y$ is the
inelasticity determined using the relation $y = Q^{2} / x_{\rm{Bj}} s$.
In this formula, $x_{\rm{Bj}}$ is the Bjorken scaling variable and  $s$ is the square of 
the centre-of-mass energy. The selected events were required to have a well reconstructed and isolated scattered electron. 
The pseudorapidities of the jets in the laboratory 
frame, $\eta^{\rm{jet}}_{\rm{LAB}}$, were required to satisfy 
$-1 < \eta^{\rm{jet}}_{\rm{LAB}} < 2.5$. For the measurement of the jet 
cross sections events with at least two jets with transverse energies in the 
Breit frame, $E^{\rm{jet}}_{T,\rm{B}}$, greater than 8 GeV were selected.  Additionally, the
invariant mass of the dijet system had to be greater than 20 GeV.
The latter cut was introduced to suppress infrared sensitive regions in the fixed-order
calculations.

\begin{figure}[h]
\parbox[c]{5.5cm}{\epsfxsize=45mm
\epsffile{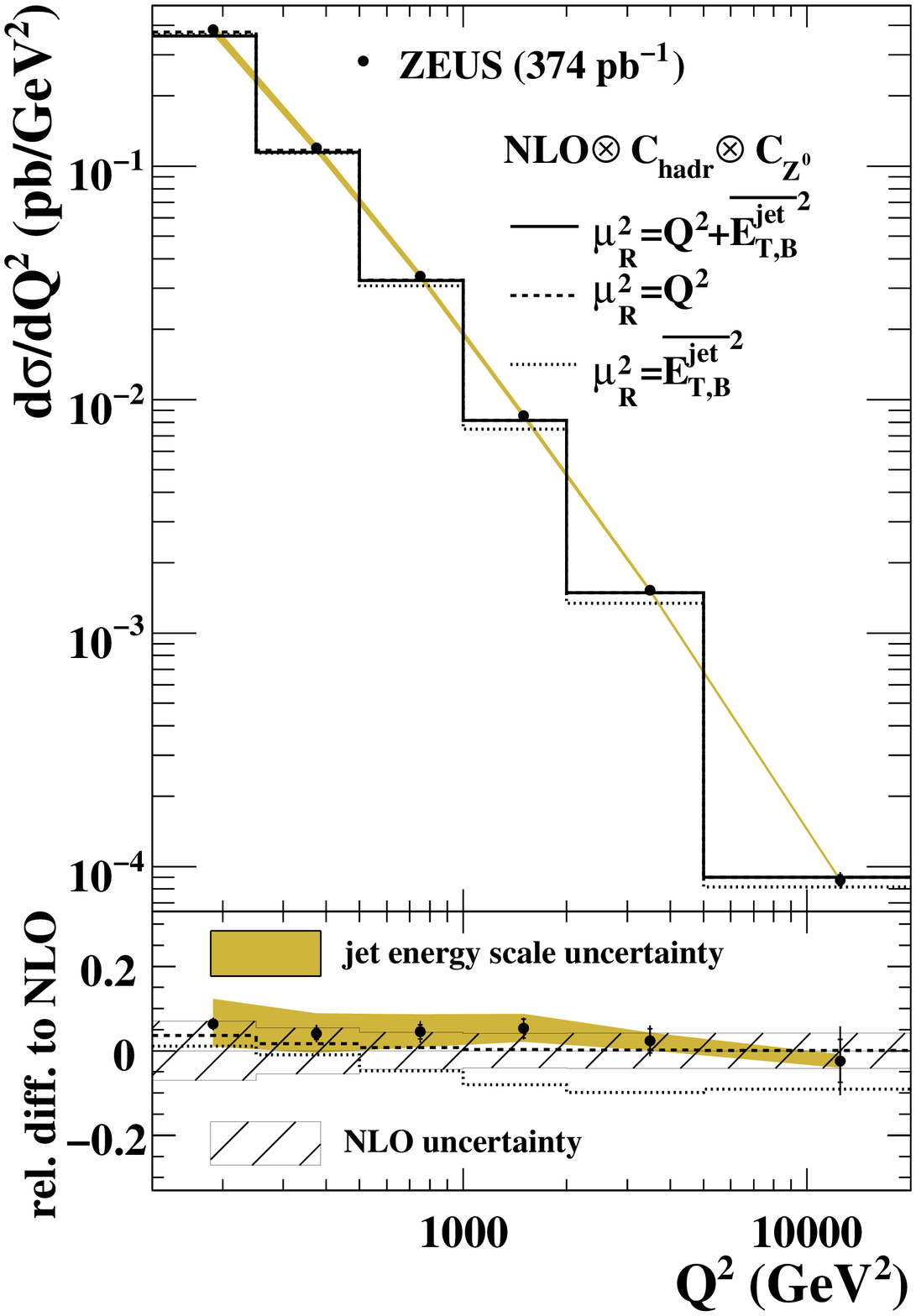}} \hfill~\parbox[c]{5.5cm}{\epsfxsize=45mm
\epsffile{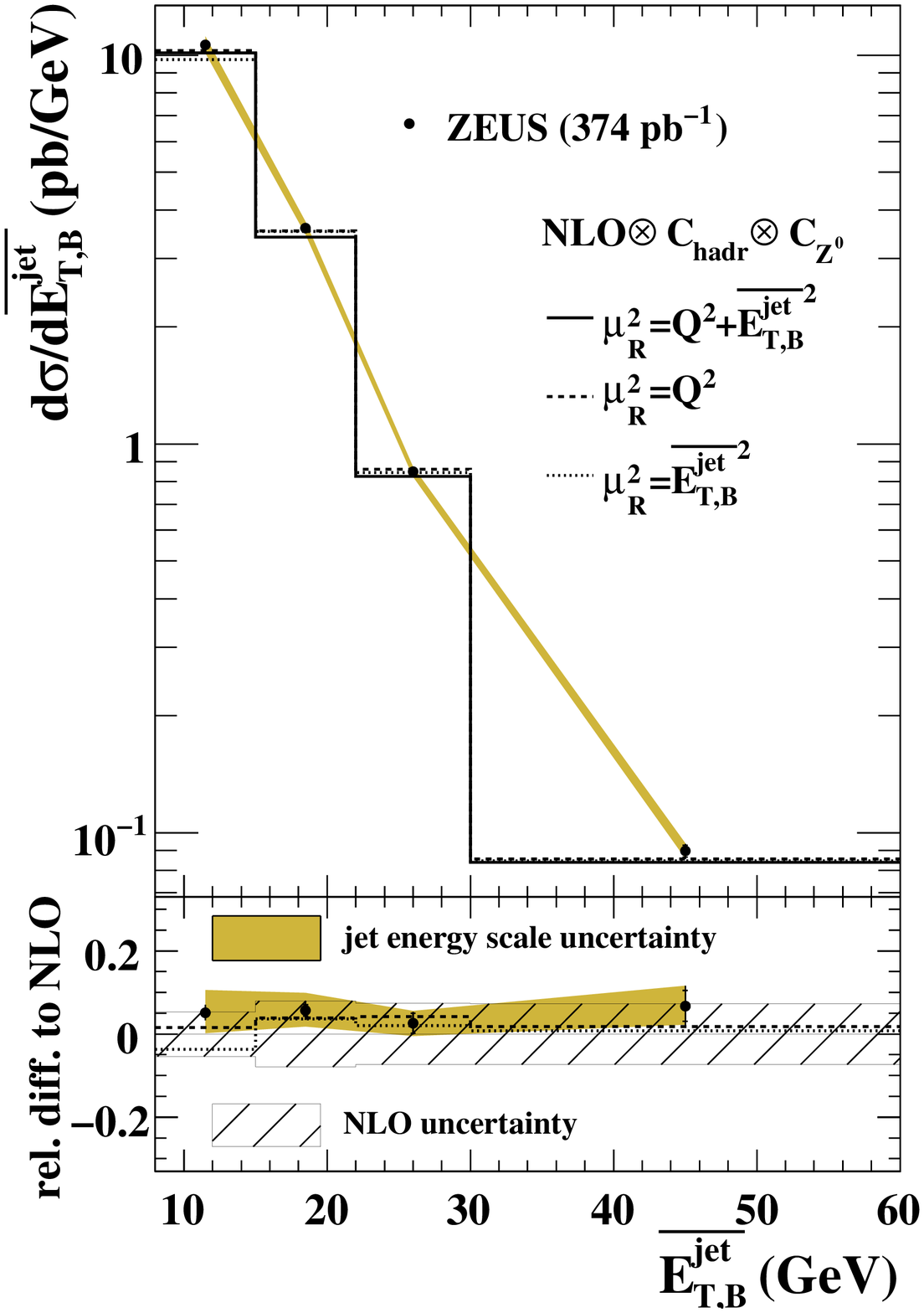}} 
\caption{The differential cross sections as functions of the exchanged boson
virtuality, $Q^{2}$ (left), and the mean energy of the jets of the 
dijet system in the Breit frame, $\overline{E^{jet}_{T, \rm{B}}}$ (right).}
\label{fig:DIS_cross_sec}
\end{figure}


Cross sections were compared to NLO QCD calculations as implemented in the
{\sc Nlojet++} program \cite{Nlojet}. The calculations were obtained using the
CTEQ6.6 parameterisations for the proton PDFs with the factorisation and 
renormalisation scales set to $\mu_{F} = Q$ and 
$\mu_{R}^{2} = Q^{2} + \overline{E_{T,\rm{B}}^{\rm{jet}}}^{2}$, respectively. Here, 
$\overline{E_{T,\rm{B}}^{\rm{jet}}}$ is the mean jet transverse energy of the 
dijet system in the Breit frame. The uncertainty on the 
NLO predictions due to missing higher orders was estimated by varying $\mu_{R}$
by a factor 2 up and down and was found to be below $\pm 6\%$ at low $Q^{2}$
and low $\overline{E^{\rm{jet}}_{T, \rm{B}}}$ and below $\pm 3\%$ in the 
highest $Q^{2}$ region. The calculations provide a good description of the 
measured cross sections, as demonstrated in Fig. \ref{fig:DIS_cross_sec}, 
where the differential cross sections as functions of $Q^{2}$ and 
$\overline{E^{\rm{jet}}_{T,\rm{B}}}$  are compared with the NLO QCD predictions.
The data and the theory agree very well in shape and normalisation. The 
theoretical uncertainty in the lower $Q^{2}$ region, which amounts to about $\pm7\%$, is larger 
than the uncertainty of the data, which is $\approx\pm1\%$. The fraction of gluon induced events as predicted by CTEQ6.6 ranges from about 75$\%$ at 
$125 < Q^{2} < 250\;\mbox{GeV}^{2}$ to about 5$\%$ in the highest $Q^{2}$ region investigated. The 
PDF uncertainty in the medium $Q^{2}$ region is larger than the theoretical 
uncertainty due to the choice of $\mu_{R}$. Therefore precise input for the  determination of the
gluon distribution function is expected.

\section{Inclusive dijets in photoproduction}

In photoproduction the electron escapes undetected through the beam pipe. Thus, the jets 
of the dijet system are approximately balancing each other in the transverse plane. 
The phase space of the measurement was defined by $Q^{2} < 1\;\mbox{GeV}^{2}$ 
with the centre-of-mass energy of the photon-proton system, $W_{\gamma p}$, in the range 
$142 < W_{\gamma p} < 293\;\mbox{GeV}$. Selected events were required to lack an 
identified scattered electron.
For the cross sections presented here, only events with at least two jets with 
transverse energies $E^{\rm{jet}1}_{T} > 21\;\mbox{GeV}$ and 
$E^{\rm{jet}2}_{T} > 17\;\mbox{GeV}$ were considered. The latter cuts, which are
asymmetric, were applied to make the theory infrared insensitive.
The pseudorapidities of the jets had to satisfy $-1 < \eta^{\rm{jet}} < 2.5$ to 
ensure that jets were contained within a section of the detector in which the 
acceptance is well understood.


In order to distinguish direct and resolved photoproduction events, the variable
$x_{\gamma}$ was used, which is the fraction of the photon momentum 
participating in the production of the two most energetic jets. This variable 
can be determined according to $x^{\rm{obs}}_{\gamma} = 
(E^{\rm{jet}1}_{T} e^{-\eta^{\rm{jet}1}} + E^{\rm{jet}2}_{T} e^{-\eta^{\rm{jet}2}})/{2yE_{e}}$,
where $E_{e}=27.5\;\rm{GeV}$ is the energy of the electron beam. For 
direct photon events, $x_{\gamma}$ is close to one, whereas for events with resolved 
photons characteristic $x_{\gamma}$ values are smaller.

\begin{figure}[h]
\parbox[c]{6.cm}{\epsfxsize=60mm
\epsffile{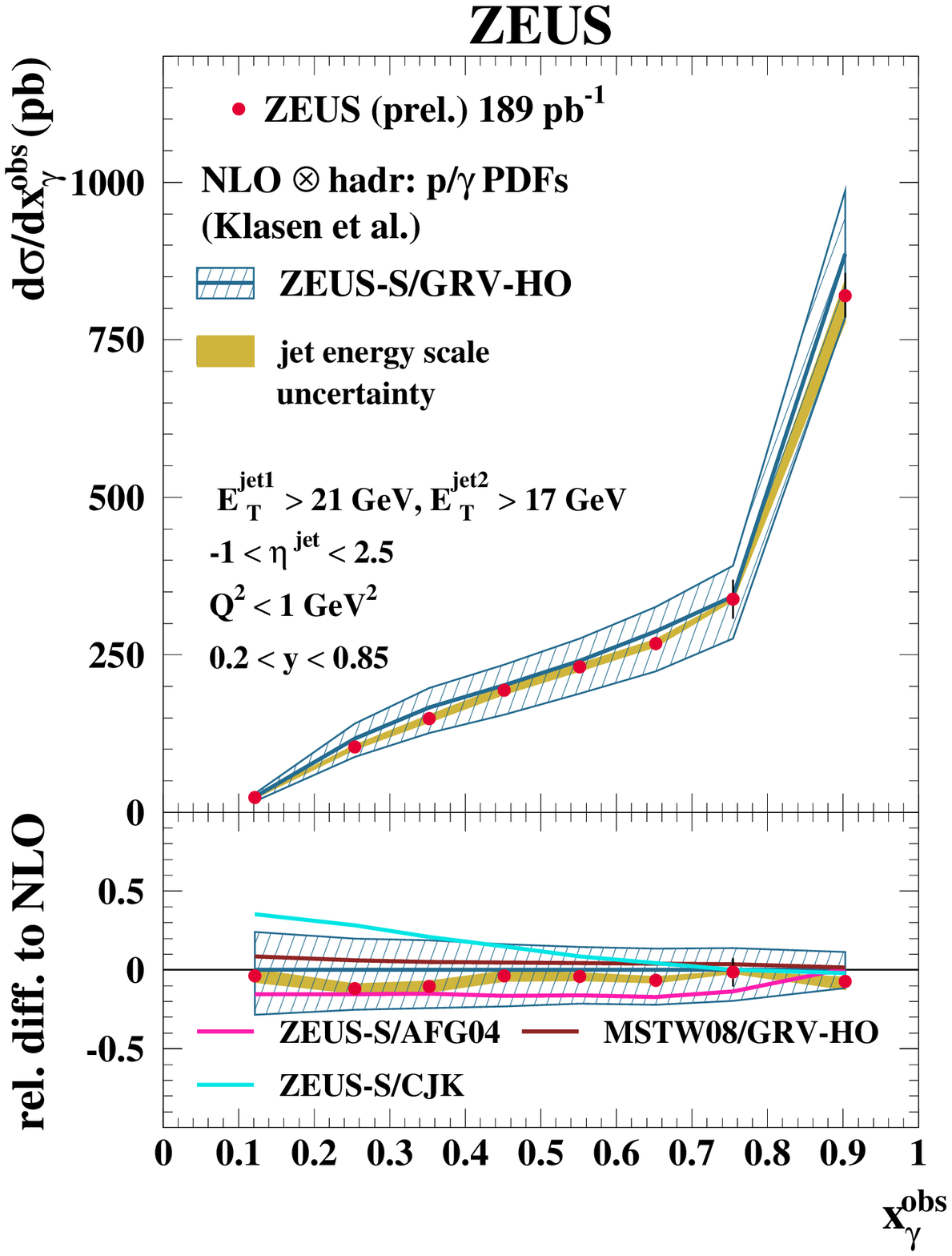}} \hfill~\parbox[c]{6.cm}{\epsfxsize=60mm
\epsffile{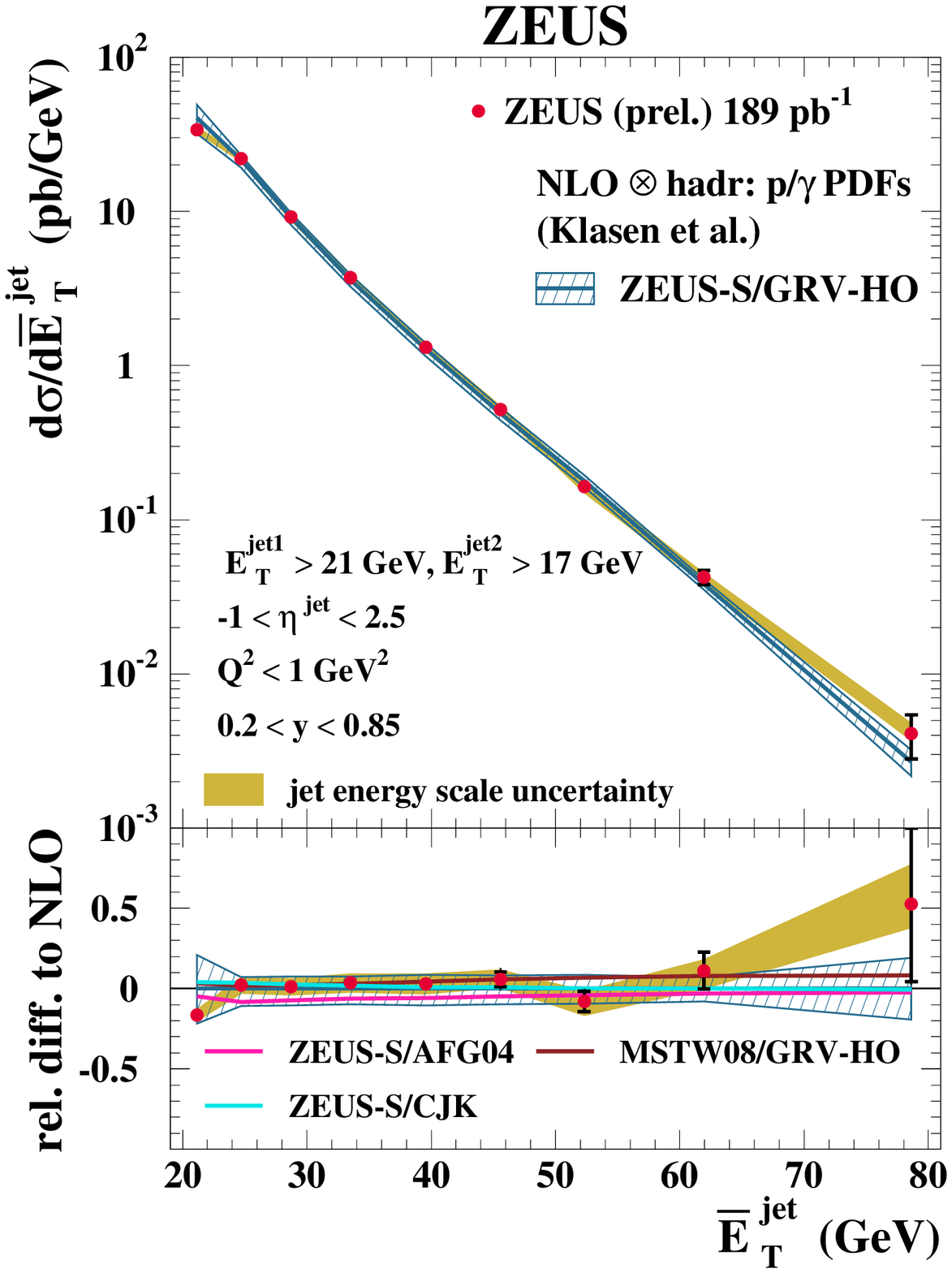}} 
\caption{The differential cross sections as functions of $x^{\rm{obs}}_{\gamma}$ 
(left) $\overline{E^{\rm{jet}}_{T}}$ (right). \label{fig:PHP_cross_sec}}
\end{figure}

Cross sections were compared to NLO QCD calculations obtained using
the program from Klasen, Kleinwort and Kramer \cite{Klasen}. The ZEUS-S proton PDFs and the GRV-HO
 photon PDFs were used. The scales $\mu_{R}$ and  $\mu_{F}$ were set to 
$\mu_{R} = \mu_{F} = (E^{\rm{jet}}_{T})^{\rm{max}}$. The renormalisation scale
uncertainty was evaluated by scaling $\mu_{R}$ by a factor 2 up and down and amounts to 
20$\%$. As shown in Fig. \ref{fig:PHP_cross_sec}, where the cross 
sections as functions of $x^{\rm{obs}}_{\gamma}$ and $\overline{E^{\rm{jet}}_{T}}$  are compared with NLO QCD predictions, the NLO calculations provide
a good description of the data. 
The cross section measured as a function of $x^{\rm{obs}}_{\gamma}$ is sensitive 
to the photon PDFs, due to the large spread observed between 
the predictions using different parameterisations of the photon PDFs. This
sensitivity is especially pronounced in the low-$x^{\rm{obs}}_{\gamma}$ region, 
in which resolved photon events are dominating. In all investigated cross section bins,
the theoretical uncertainties are larger than the experimental uncertainties.

The measured differential cross sections of inclusive dijet production in both 
neutral current DIS and photoproduction have small statistical and 
systematic uncertainties. The description of the data by NLO QCD is
good. These jet
data have the potential to significantly reduce PDF uncertainties and provide
information for the determination of the strong coupling constant.

\bibliographystyle{aipproc}

\vfill \eject
\end{document}